\documentclass[preprint,12pt]{elsarticle}
\journal{International Journal of Engineering Science}
\usepackage{graphics,epsfig}
\usepackage{epstopdf}
\usepackage{float}
\usepackage{amssymb,amstext,amsmath}
\usepackage{mathtools}
\usepackage{booktabs}
\usepackage{natbib}
\setcitestyle{round,aysep={,},yysep={,},citesep={;}}
\usepackage{physics}

\begin{document}
\newcommand{\balpha}{\boldsymbol{\alpha}}
\newcommand{\bbeta}{\boldsymbol{\beta}}
\newcommand{\bgamma}{\boldsymbol{\gamma}}
\newcommand{\bdelta}{\boldsymbol{\delta}}
\newcommand{\bepsilon}{\boldsymbol{\epsilon}}
\newcommand{\bvarepsilon}{\boldsymbol{\varepsilon}}
\newcommand{\bzeta}{\boldsymbol{\zeta}}
\newcommand{\bfoldeta}{\boldsymbol{\eta}}
\newcommand{\btheta}{\boldsymbol{\theta}}
\newcommand{\bvartheta}{\boldsymbol{\vartheta}}
\newcommand{\biota}{\boldsymbol{\iota}}
\newcommand{\bkappa}{\boldsymbol{\kappa}}
\newcommand{\blambda}{\boldsymbol{\lambda}}
\newcommand{\bmu}{\boldsymbol{\mu}}
\newcommand{\bnu}{\boldsymbol{\nu}}
\newcommand{\bxi}{\boldsymbol{\xi}}
\newcommand{\bpi}{\boldsymbol{\pi}}
\newcommand{\bvarpi}{\boldsymbol{\varpi}}
\newcommand{\brho}{\boldsymbol{\rho}}
\newcommand{\bvarrho}{\boldsymbol{\varrho}}
\newcommand{\bsigma}{\boldsymbol{\sigma}}
\newcommand{\bvarsigma}{\boldsymbol{\varsigma}}
\newcommand{\btau}{\boldsymbol{\tau}}
\newcommand{\bupsilon}{\boldsymbol{\upsilon}}
\newcommand{\bphi}{\boldsymbol{\phi}}
\newcommand{\bvarphi}{\boldsymbol{\varphi}}
\newcommand{\bchi}{\boldsymbol{\chi}}
\newcommand{\bpsi}{\boldsymbol{\psi}}
\newcommand{\bomega}{\boldsymbol{\omega}}
\newcommand{\bGamma}{\boldsymbol{\Gamma}}
\newcommand{\bDelta}{\boldsymbol{\Delta}}
\newcommand{\bTheta}{\boldsymbol{\Theta}}
\newcommand{\bLambda}{\boldsymbol{\Lambda}}
\newcommand{\bXi}{\boldsymbol{\Xi}}
\newcommand{\bPi}{\boldsymbol{\Pi}}
\newcommand{\bSigma}{\boldsymbol{\Sigma}}
\newcommand{\bUpsilon}{\boldsymbol{\Upsilon}}
\newcommand{\bPhi}{\boldsymbol{\Phi}}
\newcommand{\bPsi}{\boldsymbol{\Psi}}
\newcommand{\bOmega}{\boldsymbol{\Omega}}
\newcommand{\llbracket}{[\![}
\newcommand{\rrbracket}{]\!]}
\def\Xint#1{\mathchoice
   {\XXint\displaystyle\textstyle{#1}}%
   {\XXint\textstyle\scriptstyle{#1}}%
   {\XXint\scriptstyle\scriptscriptstyle{#1}}%
   {\XXint\scriptscriptstyle\scriptscriptstyle{#1}}%
   \!\int}
\def\XXint#1#2#3{{\setbox0=\hbox{$#1{#2#3}{\int}$}
     \vcenter{\hbox{$#2#3$}}\kern-.5\wd0}}
\def\ddashint{\Xint=}
\def\dashint{\Xint-}

\begin{frontmatter}
\title{Averaging in dislocation mediated plasticity}
\author{K. C. Le$^{a,b}$\footnote{Corresponding author: lekhanhchau@tdtu.edu.vn.}, T.H. Le$^{c}$, T. M. Tran$^{a,b}$}
\address{$^a$Materials Mechanics Research Group, Ton Duc Thang University, Ho Chi Minh City, Vietnam
\\
$^b$Faculty of Civil Engineering, Ton Duc Thang University, Ho Chi Minh City, Vietnam
\\
$^c$Computational Engineering, Vietnamese German University, Binh Duong, Vietnam}
\begin{abstract} 
Starting from the assumption that all possible orientations of grains are equally probable, we evaluate the mean resolved shear stress and strains in polycrystalline bars subjected to axially symmetric tension or compression. The evaluated quantities, in combination with the kinetics of thermally activated dislocation depinning, are used to establish the equation for the stress rate. We then perform a large-scale least-square analysis to identify the physics based parameters of this theory and show that the simulated stress-strain curves for copper, iron and steel agree well with the experiments of Johnson and Cook. 
\end{abstract}

\end{frontmatter}

\section{Introduction}
The simplest version of thermodynamic dislocation theory (TDT), initiated by \citet{LBL-10}, deals with single crystals deforming in simple shear. To apply this theory to tension/compression tests, the so-called ``geometric factor'' $\alpha$ must be introduced into the stress-strain relation. Assuming the plane strain deformation of a single crystal whose slip system is inclined at an angle of 45$^\circ$ to the bar axis, we have shown in \citep{LTL-17} that $\alpha=4$. However, this result is not valid for polycrystalline bars under axially symmetric tension/compression tests, so that in all numerical simulations in \citep{LBL-10,LTL-17} the geometric factor $\alpha=1$ was assumed. In \citep{LB-18} Hooke's tensorial law is used in rate form together with other equations of TDT, which allow only a complicated and elusive numerical treatment. The aim of this paper is to develop the thermodynamic dislocation theory for polycrystalline bars under axially symmetric tension/compression by averaging the resolved shear stress and strains over all possible grain orientations. Although the macroscopic stress state of the stretched (compressed) bar is uniform, the microscopically resolved shear stress in its polycrystalline material, which is responsible for the local plastic slip in the grains, is non-uniform as it depends on the orientation of the slip systems. Our main assumption is that in any representative volume element of the polycrystal containing a large number of randomly oriented grains, all possible grain orientations are equally probable. Consider a slip system, say one from those 12 slip systems in fcc crystal. Since the orientation of this slip system in polycrystalline material is determined by the grain orientation, all its orientations are also equally probable. Under this assumption it is possible to calculate the mean resolved shear stress and the mean plastic and elastic resolved shear strain over all orientations of this slip system as well as over all slip systems. Let us mention that this idea of averaging stresses and strains over all orientations of slips is not new: It goes back to the classical paper by \citet{BB-49} within phenomenological plasticity without using dislocations. With these quantities and the homogenization technique, we can establish the stress-strain relation for a stretched (compressed) polycrystalline bar and show that the geometric factor must be $2(1+\nu)$, where $\nu$ is Poisson's ratio. More importantly, we are able to use the mean plastic resolved shear strain in combination with the kinetics of thermally activated dislocation depinning \citep{LBL-10} to derive the equation for the stress rate in polycrystalline bars under tension/compression. Based on the obtained system of equations of TDT, we perform a large-scale least squares analysis to identify the physics-based parameters of this theory and use them to predict the mechanical responses of OFHC copper, ARMCO iron and 4340 steel reported in \citep{JC-85}.
\section{Mean resolved shear stress and strains}
\begin{figure}[htb]
\centering \includegraphics[height=7cm]{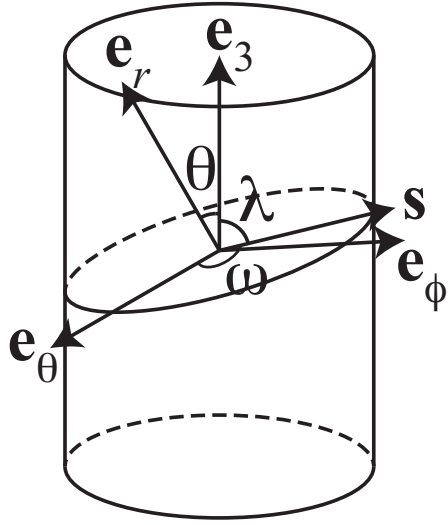} \caption{Basis vectors and slip direction} \label{fig:1}
\end{figure}

To find the mean resolved shear stress and strains, it is convenient to use the orthonormal basis vectors of the spherical coordinates
\begin{align*}
\vb{e}_r&=(\sin \theta \cos \phi , \sin \theta \sin \phi , \cos \theta)^T, \notag
\\
\vb{e}_\phi &=(\cos \theta \cos \phi , \cos \theta \sin \phi , -\sin \theta)^T, 
\\
\vb{e}_\theta &=(-\sin \phi , \cos \phi , 0)^T. \notag
\end{align*} 
The orientation of a slip system in polycrystalline material is uniquely defined by vector $\vb{m}=\vb{e}_r$, being the unit normal to the slip plane, and vector
\begin{equation*}
\vb{s}=\cos \omega \, \vb{e}_\phi +\sin \omega \, \vb{e}_\theta 
\end{equation*}
denoting the slip direction (see Fig.~\ref{fig:1}). Neglecting the stress fluctuation in the grains that does not contribute to the mean resolved shear stress, let us evaluate the latter from the macroscopic stress. For the uniform and uni-axial macroscopic stress state $\bsigma =\text{diag} (0,0,\sigma )$ the resolved shear stress (Schmid stress) caused by it in the grain with the slip system orientation $(\vb{s},\vb{m})$ equals
\begin{equation}
\label{eq:3}
\tau (\theta ,\omega )=\vb{s}\vdot \bsigma \vdot \vb{m}=-\sigma \cos \theta \sin \theta \cos \omega .
\end{equation}
Note that Eq.~\eqref{eq:3} reduces to the formula $\tau =\sigma \cos \theta \cos \lambda $, with $\cos \theta \cos \lambda $ being the Schmid factor, by observing that $\cos \lambda =\vb{s}\vdot \vb{e}_3=-\sin \theta \cos \omega $.

%%%%%%%%%%%%% FIGURE 1 %%%%%%%%%%%%
%%%%%%%%%%%%%%%%%%%%%%%%%%%%%%%

For each orientation $(\vb{s},\vb{m})$ there are the associated orientations $(-\vb{s},\vb{m})$ and $(\vb{s},-\vb{m})$, whose resolved shear stress is opposite and responsible for the plastic slip in the opposite direction. In order to find the physically meaningful average resolved shear stress, which is responsible for the positive slip direction only, we have to exclude these orientations. For this purpose we limit the angle $\theta $ to change from 0 to $\pi/2$, the angle $\omega$ to change from $\pi/2$ to $3\pi/2$, while the angle $\phi$ can change from 0 to $2\pi$. Using the equal probability hypothesis, we take the probability measure of all positive orientations as  
$$\frac{\dd \omega}{\pi }\frac{\dd a}{2\pi }=\frac{\dd \omega}{\pi}\frac{\sin \theta \dd \phi \dd \theta }{2\pi },$$
with denominators being the total length (area) of the domains of $\omega$ and $(\phi ,\theta )$. Now the mean resolved shear stress becomes
\begin{equation}
\label{eq:4}
\bar{\tau}=\int_0^{\pi/2}\int_{0}^{2\pi}\int_{\pi/2}^{3\pi/2} \sin \theta \, \tau (\theta ,\omega )\frac{\dd{\omega}}{\pi} \frac{\dd{\phi} \dd{\theta} }{2\pi }=\frac{2\sigma }{3\pi }.
\end{equation}

Likewise we can also find the mean plastic resolved shear strain from the macroscopic plastic strain. Under the axially symmetric tension/compression test the macroscopic plastic strain tensor in the bar turns out uniform and diagonal: $\bvarepsilon^p=\text{diag} (\varepsilon^p_1,\varepsilon^p_2,\varepsilon^p_3)$. Neglecting the plastic strain fluctuation, we find that the plastic resolved shear strain in the grain with  the slip system orientation $(\vb{s},\vb{m})$ caused by $\bvarepsilon^p$ is given by
\begin{multline*}
\varepsilon^p_{sm}(\phi ,\theta ,\omega )=\vb{s}\vdot \bvarepsilon^p \vdot \vb{m}
=\sin \theta  [ (-\varepsilon^p_1+\varepsilon^p_2)\sin \omega \cos \phi \sin \phi 
\\
 +\cos \omega \cos \theta (-\varepsilon^p_3 +\varepsilon^p_1 \cos^2 \phi +\varepsilon^p_2 \sin^2 \phi )] .
\end{multline*}
Thus, doing the averaging over all positive orientations, we obtain
\begin{align*}
\bar{\varepsilon}^p_{sm}&=\int_0^{\pi/2}\int_{0}^{2\pi}\int_{\pi/2}^{3\pi/2} \sin \theta \, \varepsilon^p_{sm} (\phi, \theta ,\omega )\frac{\dd{\omega}}{\pi} \frac{\dd{\phi} \dd{\theta} }{2\pi }
\\
&=\frac{3\varepsilon^p_3-(\varepsilon^p_1+\varepsilon^p_2+\varepsilon^p_3)}{3\pi }.
\end{align*}
Taking into account the incompressibility condition, $\varepsilon^p_1+\varepsilon^p_2+\varepsilon^p_3=0$, we get 
\begin{equation*}
\bar{\varepsilon}^p_{sm}=\frac{\varepsilon^p_3}{\pi }.
\end{equation*}

We turn now to the elastic resolved shear strain, $\varepsilon^e_{sm}=\vb{s}\vdot (\bvarepsilon-\bvarepsilon^p) \vdot \vb{m}$. Provided the macroscopic elastic strain tensor is uniform and diagonal, $\bvarepsilon^e=\text{diag} (\varepsilon^e_1,\varepsilon^e_2,\varepsilon^e_3)$, the same averaging procedure yields
\begin{equation*}
\bar{\varepsilon}^e_{sm}=\frac{3\varepsilon^e_3-(\varepsilon^e_1+\varepsilon^e_2+\varepsilon^e_3)}{3\pi }.
\end{equation*}
However, in contrast to the incompressible plastic deformation, the volumetric change due to the macroscopic elastic deformation, $\varepsilon^e_1+\varepsilon^e_2+\varepsilon^e_3$, does not vanish. The latter can be found by using the traction-free boundary condition at the side boundary of the bar yielding $\varepsilon^e_1=\varepsilon^e_2=-\nu \varepsilon^e_3$. Plugging these elastic strains into the previous equation we obtain  
\begin{equation}
\label{eq:8}
\bar{\varepsilon}^e_{sm}=\frac{2(1+\nu)\varepsilon^e_3}{3\pi }.
\end{equation}
From this averaging procedure we see that all slip systems behave in the same way and formulas \eqref{eq:4} and \eqref{eq:8} remain valid for the mean resolved shear stress and strains over all positive orientations and all slip systems.

Applying homogenization technique to heterogeneous materials with statistically isotropic microstructure \citep{Torquato2002} we can establish the following relation
\begin{equation*}
\bar{\tau}=2\mu \bar{\varepsilon}^e_{sm},
\end{equation*}
where $\mu $ is the effective shear modulus that may depend on temperature. Dropping index 3 of $\varepsilon_3$ and $\varepsilon^p_3$ for short and substituting Eqs.~\eqref{eq:4} and \eqref{eq:8} into the previous equation, we finally arrive at
\begin{equation}
\label{eq:10}
\sigma =2\mu (1+\nu)(\varepsilon-\varepsilon^p),
\end{equation}
which shows that the ``geometric factor'' for polycrystalline bar under tension/compression is $\alpha=2(1+\nu)$.

\section{Equation of motion}
We further assume that the mean plastic slip rate can be found from the kinetics of thermally activated dislocation depinning as follows
\begin{equation*}
\dot{\bar{\gamma}}^pt_0=b\sqrt{\rho }\exp \Bigl( -\frac{T_p}{T}e^{-\bar{\tau }/\tau_T} \Bigr) .
\end{equation*} 
Here, $t_0$ is a microscopic time of the order of $10^{-12}$s, $b$ the Burgers vector, $\rho$ the dislocation density, $\theta=k_B T$ the ordinary temperature (in the energy unit), $k_B T_P$ the pinning energy barrier at zero stress, and $\tau_T= \mu_T b\sqrt{\rho}$ the Taylor stress \citep{LBL-10}. Since $\bar{\tau}=2\sigma/3\pi$ and $\dot{\bar{\gamma}}^pt_0=2\dot{\bar{\varepsilon}}_{sm}^pt_0=\frac{2}{\pi}\dot{\varepsilon}^pt_0$ we have
\begin{equation}
\label{eq:12}
\dot{\varepsilon}^pt_0=\frac{\pi}{2}q(\sigma,\rho,T), 
\end{equation}
where
\begin{displaymath}
q(\sigma,\rho,T)=b\sqrt{\rho }\exp \Bigl( -\frac{T_p}{T}e^{-\sigma /\bar{\mu}_Tb\sqrt{\rho}} \Bigr) ,
\end{displaymath}
and $\bar{\mu}_T=3\pi \mu_T/2$. Taking the time derivative of Eq.~\eqref{eq:10} we obtain
\begin{equation}
\label{eq:13}
\dot{\sigma}=2\mu (1+\nu)(\dot{\varepsilon}-\dot{\varepsilon}^p).
\end{equation}
For the constant strain rate $\dot{\varepsilon}=q_0/t_0$ we can replace the time derivative by the derivative with respect to the strain according to $\dv*{}{t}=(q_0/t_0)\dv*{}{\varepsilon}$. Then Eq.~\eqref{eq:13}, together with \eqref{eq:12}, is transformed to
\begin{equation}
\label{eq:14}
\dv{\sigma}{\varepsilon}=2\mu (1+\nu)\Bigl( 1-\frac{q}{\bar{q}_0}\Bigr),
\end{equation} 
where $\bar{q}_0=2q_0/\pi$.

The next logical step is to derive the equation for the dislocation density $\rho$. To this end we adopt the main hypothesis of TDT formulated first by \citet{LBL-10}: A plastically deforming crystal, as thermodynamic system, can be decomposed into two subsystems. The first, configurational subsystem, consists of slowly changing stable positions of atoms, including the positions of dislocations. The second, kinetic-vibrational  subsystem, consists of rapidly changing coordinates that describe small oscillations of atoms about their stable positions. Because of the different time scales characterizing these subsystems we can introduce two well-defined thermodynamic temperatures: the effective disorder temperature induced by dislocations, $\chi$, that will be the essential state variable, and the ordinary temperature, $\theta$, which we will assume to be a constant. Now the equations of motion for $\rho$ and $\chi$ can be derived from the non-equilibrium thermodynamics of driven systems \citep{LBL-10}. Without going into the detailed derivation, we write down these equations:
\begin{align}
\dv{\rho}{\varepsilon}&=\frac{\kappa_\rho }{\bar{\mu}_T \zeta ^2(\rho,\bar{q}_0,T)b^2} \frac{\sigma q}{\bar{q}_0}\Bigl[1-\frac{\rho}{\rho_s(\chi)}\Bigr] , \label{eq:15}
\\
\dv{\chi}{\varepsilon}&=\frac{\kappa_\chi }{\bar{\mu}_T}\frac{\sigma q}{\bar{q}_0} \Bigl(1-\frac{\chi}{\chi_0}\Bigr) . \notag
\end{align}
Here $\rho_{s}(\chi)=\frac{1}{a^2}e^{-e_d/\chi }$ is the most probable (steady-state) dislocation density at fixed configurational temperature $\chi$, while
\begin{displaymath}
\zeta(\rho,\bar{q}_0,T)=\ln \Bigl( \frac{T_P}{T}\Bigr)-\ln\Bigl[\ln\Bigl(\frac{b\sqrt{\rho}}{\bar{q}_0}\Bigr)\Bigr] .
\end{displaymath}

It is convenient to introduce the rescaled variables and rewrite the system \eqref{eq:14} and \eqref{eq:15} in the dimensionless form. First, let us introduce the dimensionless dislocation density and effective disorder temperature as
\begin{equation*}
\tilde{\rho }=a^2 \rho, \quad \tilde{\chi }=\frac{\chi }{e_d}.
\end{equation*}
With these rescaled quantities we find the dimensionless steady-state dislocation density 
\begin{equation*}
\tilde{\rho}_s=e^{-1/\tilde{\chi }}.
\end{equation*}
We introduce the dimensionless ordinary temperature $\tilde{\theta} =T/T_P$. Then the normalized plastic strain rate becomes
\begin{equation*}
q(\sigma,\rho,T)=\frac{2}{\pi }\dot{\bar{\varepsilon}}^pt_0=(b/a)\tilde{q}(\sigma,\tilde{\rho},\tilde{\theta}),
\end{equation*}
where
\begin{equation*}
\tilde{q}(\sigma,\tilde{\rho},\tilde{\theta})=\sqrt{\tilde{\rho}}\exp \Bigl[-\frac{1}{\tilde{\theta}} e^{-\sigma/\tilde{\mu}_T\sqrt{\tilde{\rho}}} \Bigr] ,\quad \tilde{\mu}_T=(b/a)\bar{\mu}_T.
\end{equation*}
The formula for $\zeta$ becomes
\begin{equation*}
\tilde{\zeta}(\tilde{\rho},\tilde{q}_0,\tilde{\theta})=\ln \Bigl( \frac{1}{\tilde{\theta}}\Bigr)-\ln\Bigl[\ln\Bigl(\frac{\sqrt{\tilde{\rho}}}{\tilde{q}_0}\Bigr)\Bigr] . 
\end{equation*}
We assume that $\tilde{\mu}_T$ scales like $\mu$ as a function of temperature $\tilde{\mu}_T(\tilde{\theta})=s\mu(\tilde{\theta})$, with $s$ being a material constant. Using $\tilde{q}$ instead of $q$ as the dimensionless measure of mean plastic slip rate, we are effectively rescaling $t_0$ by a factor $b/a$. For the purpose of this analysis, we assume that $(2a/\pi b)t_0=10^{-12}$s; and we use $\tilde{q}_0=10^{-12}\dot{\varepsilon}$ for converting from $\tilde{q}_0$ to the measured total strain rates.

In terms of the introduced rescaled variables the governing equations read
\begin{align}
\dv{\sigma}{\varepsilon}&=2\mu (1+\nu) \Bigl( 1-\frac{\tilde{q}}{\tilde{q}_0}\Bigr), \notag
\\
\dv{\tilde{\rho}}{\varepsilon}&=K_\rho \frac{\sigma }{\tilde{\mu }_T \tilde{\zeta}^2(\tilde{\rho},\tilde{q}_0,\tilde{\theta})}\frac{\tilde{q}}{\tilde{q}_0}\Bigl[1-\frac{\tilde{\rho}}{\tilde{\rho}_s(\tilde{\chi})}\Bigr] , \label{eq:16}
\\
\dv{\tilde{\chi}}{\varepsilon}&=K_\chi \frac{\sigma }{\tilde{\mu}_T}\frac{\tilde{q}}{\tilde{q}_0} \Bigl(1-\frac{\tilde{\chi}}{\tilde{\chi}_0}\Bigr) ,\notag
\end{align}
where $K_\rho =\frac{\kappa_\rho a}{b}$ and $K_\chi = \frac{\kappa_\chi b}{a e_d}$. We  assume that the coefficients $K_\rho $ and $K_\chi$ are material constants independent of the strain rate and temperature.

\section{Data analysis}

%%%%%%%%%%%%% FIGURE 2 %%%%%%%%%%%%
\begin{figure}[htb]
\centering \includegraphics[width=8cm]{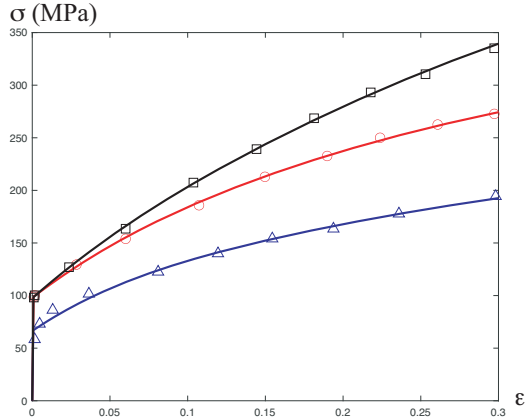} \caption{(Color online) Stress-strain curves for OFHC copper: (i) at $\dot\varepsilon = 451\,$s$^{-1}$, $T=298\,$K (black); (ii) at $\dot\varepsilon = 449\,$s$^{-1}$,  $T=496\,$K (red); (iii) at $\dot\varepsilon = 464\,$s$^{-1}$,  $T=732\,$K (blue).  The experimental points (black squares, red circles, and blue triangles) are taken from \citet{JC-85}} \label{fig:copper}
\end{figure}
%%%%%%%%%%%%%%%%%%%%%%%%%%%%%%%%%%%%% 

\begin{figure}[htb]
\centering \includegraphics[width=8cm]{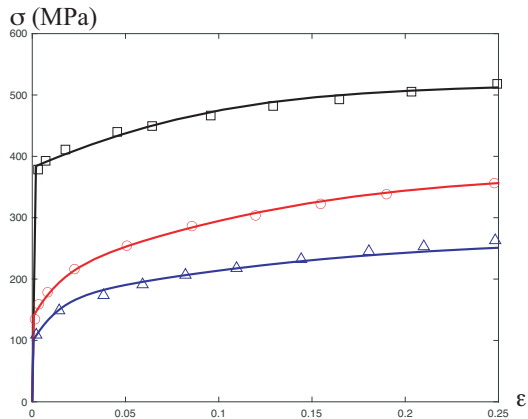} \caption{(Color online) Stress-strain curves for ARMCO iron: (i) at $\dot\varepsilon = 407\,$s$^{-1}$, $T=298\,$K (black); (ii) at $\dot\varepsilon = 435\,$s$^{-1}$,  $T=500\,$K (red); (iii) at $\dot\varepsilon = 460\,$s$^{-1}$,  $T=735\,$K (blue).  The experimental points (black squares, red circles, and blue triangles) are taken from \citet{JC-85}} \label{fig:iron}
\end{figure}

\begin{figure}[htb]
\centering \includegraphics[width=8cm]{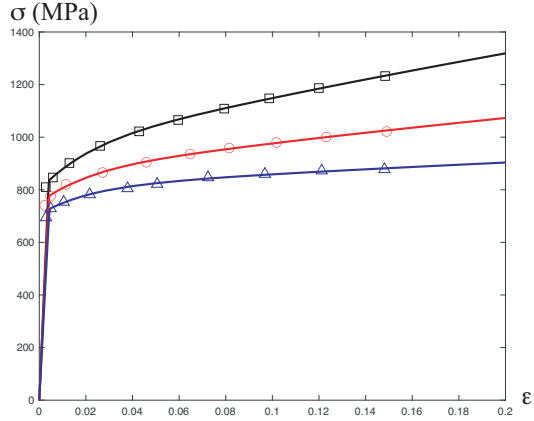} \caption{(Color online) Stress-strain curves for 4340 steel: (i) at $\dot\varepsilon = 570\,$s$^{-1}$, $T=298\,$K (black); (ii) at $\dot\varepsilon = 604\,$s$^{-1}$,  $T=500\,$K (red); (iii) at $\dot\varepsilon = 650\,$s$^{-1}$,  $T=735\,$K (blue).  The experimental points (black squares, red circles, and blue triangles) are taken from \citet{JC-85}} \label{fig:steel}
\end{figure}

The experimental results of \citet{JC-85} for OFHC copper, ARMCO iron, and 4340 steel, along with our theoretical results based on the governing equations \eqref{eq:16}, are shown in Figs. \ref{fig:copper}, \ref{fig:iron}, and \ref{fig:steel}, respectively. Note that the strain, defined by \citet{JC-85} as $2\ln (d_0/d)$, approximately equals $\varepsilon_3=\varepsilon$, provided the small strain and incompressibility conditions are satisfied. In order to compute the theoretical curves in these figures, we need values for five system-specific parameters: the activation temperature $T_P$, the stress ratio $s$, the steady-state scaled effective temperature $\tilde\chi_0$, and the two dimensionless conversion factors $K_\rho$ and $K_\chi$. We also need initial values of the scaled dislocation density $\tilde\rho(\varepsilon = 0) \equiv \tilde\rho_i$ and the effective temperature $\tilde\chi(\varepsilon = 0) \equiv \tilde\chi_i$, which are determined by sample preparation.  Further, we need a formula for the temperature dependent shear modulus $\mu(\tilde{\theta})$, which we take from \citep{VARSHNI-70} to be 
\begin{equation*}
\mu(\tilde\theta) = \mu_1 - \frac{D}{\exp(T_1/T_P\,\tilde\theta)-1},
\end{equation*}
where $\mu_1 = 51.3\,$GPa, $D = 3\,$GPa, $T_1 = 165\,$K for copper, and $\mu_1 = 85\,$GPa, $D = 10\,$GPa, $T_1 = 298\,$K for steel. For iron we use a simple linear approximation $\mu (\tilde{\theta})=85.335 \text{GPa}-0.034 (T_P \tilde{\theta})$. Finally, Poisson's ratio is: $\nu=0.34$ (copper), $\nu=0.29$ (iron and steel) \citep{JC-85}.

%%%%%%%%%%%%% FIGURE 3 %%%%%%%%%%%%
%%%%%%%%%%%%%%%%%%%%%%%%%%%%%%%%%%%%% 

To identify the parameters we perform the large-scale least-squares analysis developed by \citet{LTL-17,LE-TRAN-17}. That is, we have computed the sum of the squares of the differences between our theoretical stress-strain curves and the experimental points, and have minimized this sum in the space of the unknown parameters. The identified parameters turn out robust and are presented in Table~\ref{tab:1} for the above materials. We also found the initial conditions for OFHC copper corresponding to the above three cases: (i) $\tilde{\chi}_i = 0.166$, $\tilde{\rho}_i = 3.8\times 10^{-4}$,  (ii) $\tilde{\chi}_i = 0.176$, $\tilde{\rho}_i = 7.16\times 10^{-4}$, (iii) $\tilde{\chi}_i = 0.17$, $\tilde{\rho}_i = 6.18\times 10^{-4}$. For ARMCO iron we found that: (i) $\tilde{\chi}_i  = 0.194$, $\tilde{\rho}_i = 5\times 10^{-3}$,  (ii) $\tilde{\chi}_i = 0.175$, $\tilde{\rho}_i = 1.24\times 10^{-3}$, (iii) $\tilde{\chi}_i  = 0.18$, $\tilde{\rho}_i = 1.26\times 10^{-3}$. Finally, for 4340 steel we have: (i) $\tilde{\chi}_i = 0.207$, $\tilde{\rho}_i = 5.22\times 10^{-3}$,  (ii) $\tilde{\chi}_i = 0.221$, $\tilde{\rho}_i = 8.06\times 10^{-3}$, (iii) $\tilde{\chi}_i = 0.244$, $\tilde{\rho}_i = 1.29\times 10^{-2}$.

%%%%%%%%%%%%% FIGURE 4 %%%%%%%%%%%%
%%%%%%%%%%%%%%%%%%%%%%%%%%%%%%%%%%%%% 

\begin{table}[h]
  \centering 
\begin{tabular}{|c|c|c|c|c|c|}
\hline
Parameters & $T_P$ (K) & $s$ & $\chi_0$ & $K_\chi$ & $K_\rho$ \\ \hline
OFHC copper & $65212$ & $0.04242$ & $0.21122$ & $10.06969$ &  $1.58631$\\  
\hline
ARMCO iron & 68150 & 0.02901 & 0.21322 & 14.56249 & 8.94699 \\ 
\hline
4340 steel & 95048 & 0.05143 & 0.28248 & 2.43387 & 12.38521 \\ \hline
\end{tabular}  
  \caption{The material parameters for OFHC copper, ARMCO iron, and 4340 steel}
  \label{tab:1}
\end{table}

The agreement between theory and experiment seems to be well within the bounds of experimental uncertainties.  Even the initial yielding transitions appear to be described accurately by this dynamical theory. 
\section{Conclusion}
Based on this good agreement with the experiments and the plausible arguments provided at the beginning of the paper, we conclude that the equations of thermodynamic dislocation theory for polycrystalline bars subjected to tension/compression must be derived from the averaging of the resolved shear stress and strains over all possible grain orientations. The equal probability hypothesis plays a decisive role here. Note, however, that this hypothesis does not apply to materials with preferred grain orientation such as drawn wires or rolled sheets of metals. For such textured materials, the averaging method must be further developed.

\bigskip
\noindent {\it Acknowledgement.}

K.C. Le thanks C.K.C. Lieou for the helpful discussion.

\end{document}